\documentclass[showpacs,aps,prb,amsmath,amssymb]{revtex4}
\usepackage{graphics}
\usepackage{pstricks,pst-node}
\usepackage{amssymb}
\begin{document}
%
%%%%%%%%%%%%%%%
\title{Extended Hubbard model with renormalized Wannier wave functions in the
correlated state II: Quantum critical scaling of the
wave function near the Mott-Hubbard transition}

\author{Jozef Spa\l{}ek$^{1,2}$, Jan Kurzyk$^{3}$, Robert Podsiad\l{}y$^{1}$,
and W\l{}odzimierz W\'ojcik$^{3}$}
\affiliation{$^{1}$Marian Smoluchowski Institute of Physics, Jagiellonian University,
Reymonta 4, 30-059 Krak\'ow, Poland \\
$^{2}$Faculty of Physics and Applied Computer Science, AGH Univ. of Science and Technology,
Reymonta 19, 30-059 Krak\'ow, Poland \\
$^{3}$Institute of Physics, Cracow University of Technology, Podchor\c{a}\.zych 1, 30-084 Krak\'{o}w,
Poland}
%
%
%\date{Received: date / Revised version: 23/11/09}
\begin{abstract}
We present a model example of a quantum critical behavior of renormalized
single-particle Wannier function composed of Slater s-orbitals and represented in
an adjustable Gaussian STO-7G basis, which is
calculated for cubic lattices in the Gutzwiller correlated state near the metal-insulator
transition (MIT). The discussion is carried out within the extended Hubbard model and the method of approach proposed earlier [cf. Eur. Phys. J. B {\bf 66}, 385 (2008)]. The component atomic-wave-function size,
the Wannier function maximum,
as well as the system energy, all scale with the increasing lattice parameter $R$ as $[(R-R_{c})/R_{c}]^{s}$ with $s$ in the interval $[0.9,1.0]$. Such scaling law is interpreted as evidence of a dominant role of the interparticle Coulomb repulsion, which for $R>R_c$ is of intersite character.
Relation of the insulator-metal transition lattice-parameter value $R=R_{c}$ to the original
{\em Mott criterion\/} is also obtained. The method feasibility is tested by comparing our results with the exact
approach for the Hubbard chain, for which the Mott-Hubbard transition is absent. In view of unique features of our results, an extensive discussion in qualitative terms is also provided.
\end{abstract}
 % end of PACS codes
\maketitle

\maketitle
%\addtolength{\baselineskip}{14pt}

\section*{1. Introduction}

Metal-insulator transition (MIT) of the Mott-Hubbard type represents one of the central problems in condensed matter physics and materials science,
since it exemplifies a transition from well defined atomic (localized or confined) states to delocalized
(Bloch- or Fermi-liquid) type of states in a solid \cite{Mott} and in other systems. \cite{Greiner} The simplest model system
represents a
lattice of spin-1/2 fermions with one particle per atomic site. Mott expressed this localized-delocalized
transformation in terms of critical density $n_{C}$ of fermions (or equivalently, in terms of critical
interparticle distance $R_{c}=n_{C}^{-1/3}$, for three-dimensional system and the effective Bohr-
radius $a$ of the atomic states formed at the transition, in the form $ n_{C}^{-1/3} a\simeq
0.2-0.25$. On the other
hand, Hubbard \cite{Hubbard} and others \cite{Brinkman} formulated the criterion in the form that at the transition the
magnitude of the intraatomic
Coulomb repulsive energy among the particles ($U$) is equal to their kinetic (band) energy
characterized
by the bare bandwidth $W$, i.e. that $U\simeq W$. These two criteria are regarded as related (cf. Appendix A).

The metal-insulator transition is also well defined as a phase transformation in
thermodynamic sense with the corresponding critical points on high- and
low-temperature sides. \cite{SpalekDatta} As close to MIT the band energy (negative) is almost compensated by the repulsive
(Coulomb) interaction, the much smaller thermal and for atomic-disorder energies can drive the metallic system towards
the state with localized spins. In effect, the resultant phase diagrams contain
first-order pressure-temperature transition line with a \textit{classical critical point}
at temperature $T_{cr}>0$ and a quantum critical point at $T=0$.
The existence of the former critical points was confirmed experimentally recently \cite{Limelette} and
shown to represent properties of the Van der Waals liquid-gas critical point
for the three-dimensional case of V$_{2}$O$_{3}$:Cr.
The basic question remains under what conditions a quantum
critical point (QCP) appears also at $T_{cr}=0$ between the metallic and insulating phases, as its existence
marks explicitly the fundamental boundary between the atomic and the condensed (delocalized, deconfined) quantum states.
The appearance of such QCP is usually obscured by the presence, at low temperatures,
of antiferromagnetism. \cite{SpalekDatta,HonigSpalek} Nonetheless, the answer to the above question,
even without inclusion of magnetic ordering, would
delineate the basic characteristics of the (quantum) critical point at
the border of localization driven by the interparticle interaction. \cite{Misawa}

MIT involves the drastic change of quantum macro states from a metal to an insulator, so a natural
and not addressed, as far as we are aware of, is the question how this drastic change of the macro state is reflected in the corresponding single-particle wave-function change when approaching the
transition from either side. The parameterized-model approaches \cite{Hubbard,Brinkman,SpalekDatta}
leave the Wannier functions determining the microscopic parameters
such as the hopping integral $t$ or Hubbard interaction $U$, as fixed when crossing the transition. On the other hand, the original Mott
approach \cite{Mott} introduces intuitively the concept of an emergent atomic state
{\em at the instability\/} of the metallic state (cf. Appendix A).
Thus, the missing question is: \textit{how the electron correlations
and the single-particle states are interrelated microscopically}?
This question is particularly important, since the Fermi-liquid state near the Mott-Hubbard localization
is particularly robust and therefore the Mott-Hubbard localization
At $T>0$ has a strongly discontinuous nature \cite{Mott}, apart from the classical critical point of $T_{cr}$ the crossover behavior for the temperatures above it. Furthermore, at the transition the renormalized band and the Coulomb energies are comparable and compensate each other. \cite{SpalekDatta}

In this paper we provide a simple answer to the above question by treating electronic states close to MIT within the approach in which both the
interelectronic correlations and
the single-particle Wannier wave functions $\{ w_{i}({\bf r})\}$
are treated {\em on the same footing\/} \cite{SpalekPodsiadly}.
By this we mean that those wave functions are determined simultaneously with the interparticle correlations
determining the many-body ground state.
As a result, we obtain a {\em singular\/} behavior
of the single-particle (Wannier) wave function renormalized by the correlations. This nonanalytic
character, and concomitant with it unique scaling laws, appear only when the wave function is
determined explicitly
{\em \`{a} posteriori\/}, i.e. \textit{is readjusted in the correlated state}.
Such a simultaneous determination of the single-particle wave function and two-particle aspects of the many-particle
dynamics
is indispensable in the situation when the single-particle and interaction
is comparable or, in some cases, the latter even becomes dominant.
In this sense, our analysis complements that performed within
either LDA+DMFT \cite{Kotliar} or LDA+U \cite{Anisimov} methods, as well as the QCP study \cite{Misawa},
all of which contain parameters
characterizing electronic correlations introduced only after the LDA or similar calculations have been carried out.
Obviously, the Gutzwiller-{\em ansatz\/} used here provides an exact wave function only in the limit
of high dimensions (i.e. above the upper critical dimension, not determined as yet), but it provides at least a mean-field-like
discussion of the renormalization aspects of the single-particle
wave function near MIT, not discussed so far in the literature.
In connection with this last limitation of our method, we would like to point out, that it would be
very interesting to apply methods such as DMFT \cite{Kotliar} to the proposed optimization of the Wannier
functions in the correlated state. However, no reliable method of evaluating explicitly
the DMFT ground-state energy as a function of microscopic parameters has been proposed so far,
starting from which the explicit wave-function optimization can be undertaken.
We should also underline, that our method
incorporating an exact diagonalization
was applied earlier to both
correlated nano- \cite{Misawa} and macro- \cite{KurzykWojcik} systems. We should also point out that this method is free from the double counting of
the repulsive Coulomb interaction.
Therefore, even though the presented below
results describe a model situation and within a rather simple scheme, they can be regarded as an essential additional ingredient to be implemented
in modeling real correlated systems near MIT. As was underlined earlier \cite{KurzykWojcik},
within the present method we can follow the system evolution as a function
of interatomic distance, not only as a function of model parameters. Such circumstance allows for an explicit analysis, at least in a qualitative manner, of the system evolution as a function of external pressure.

The structure of this paper is as follows. In Section 2 we introduce briefly the method devised before. \cite{SpalekPodsiadly,KurzykWojcik,KurzykSpalek}
In Sec. 3 we introduce an original critical scaling of the single-particle wave function characteristics
and of the ground-state energy. In Sec. 4 we show that the critical scaling is absent for one- and two-dimensional lattices,
in the former case in agreement with the absence of metal-insulator transitions, as discussed rigorously by Lieb and Wu. \cite{LiebWu}
In this Section we also derive the classic Mott criterion of the metal-insulator transition within the frame
of our approach. Sec. 5 contains both a brief discussion of extension to more realistic situation and concluding remarks. In Appendix A we
provide an elementary derivation and discussion of the original Mott criterion. \cite{Mott}, whereas in Appendix B we define an adjustable Gaussian basis in which the renormalized Wannier function has been determined explicitly.

A methodological remark is in place here. In our earlier work, \cite{KurzykWojcik,KurzykSpalek} the method of approach
composed of the diagonalization in the Fock space combined with the single-particle wave-function
renormalization in the Hilbert space, has been applied to the discussion of the metallic state up to the metal-insulator transition
point. No scaling presented here was either noticed or discussed there. \cite{KurzykSpalek} Therefore, we demonstrate here, in an explicit manner, the applicability of our method when {\em crossing\/} the metal-insulator boundary of the Mott-Hubbard type. Namely, we study explicitly the gradual evolution of renormalized Wannier functions into their atomic correspondant in the large interatomic-distance limit. On the two above
aspects rests our motivation that the present paper extends essentially our earlier approach.  \cite{SpalekPodsiadly,KurzykWojcik,KurzykSpalek}

\section*{2. The method: adjustment of the wave function in the correlated state}

We start from extended Hubbard model with inclusion of intersite Coulomb interaction
and ion-ion interaction as represented by the parameterized Hamiltonian \cite{SpalekPodsiadly,KurzykWojcik,KurzykSpalek}
\begin{equation}
H\,=\,\epsilon_{a}^{eff}\,\sum_{i}\, n_{i}\,+\,
\sum_{i\neq j\sigma}\, t_{ij}\, a_{i\sigma}^{\dagger}\, a_{j\sigma}\,+\,
U\,\sum_{i}\,
n_{i\uparrow}\, n_{i\downarrow}\,
\label{eq:w1}
\end{equation}
$$
+\,\frac{1}{2}\,\sum_{i\neq j}\,
K_{ij}\,\delta n_{i}\,\delta n_{j}\,,
$$
where the first term describes the effective atomic energy with
\begin{equation}
\epsilon_{a}^{eff}\,=\,
\epsilon_{a}\,+\, \frac{1}{2\, N}\,\sum_{i\neq j}\,\left(
\, K_{ij}\,+\, 2/\, R_{ij}\,\right)\,,
\label{eq:w2}
\end{equation}
the second represents the hopping between the nearest-neighboring sites, the third
the Hubbard (intraatomic) and the fourth is a part of intersite Coulomb interaction,
with $\delta n_{i}\equiv 1-n_{i}$ being the deviation from the integer electron occupancy
$n_{i}=1$, and $K_{ij}$ - the intersite Coulomb interaction. The term $(2/R_{ij})$
in (\ref{eq:w2}) expresses the classical Coulomb repulsion (in atomic units) for two ions
separated by the distance $R_{ij}$.
It is important to note that the presence of intersite Coulomb interaction is necessary in achieving
a proper atomic limit when $R\rightarrow\infty$. The model has been defined
in details earlier. \cite{SpalekPodsiadly,Anisimov,KurzykSpalek}

In the Mott-Hubbard insulating ground-state the hole-hole correlations are absent i.e.
$\langle (1-n_{i})(1-n_{j})\rangle=0$.
The role of these correlations is indeed negligible if the fundamental correlation function $d\equiv
\left\langle  n_{i\uparrow}n_{i\downarrow}\right\rangle$ vanishes or is very small, as also discussed earlier. \cite{KurzykSpalek} It vanishes in the Mott insulating state (for $R>R_c$).

The microscopic parameters of this model are expressed via the Wannier functions in the standard manner \cite{SpalekPodsiadly,KurzykWojcik},
$\{ w_{i}({\bf r})\} \equiv \{ w({\bf r}-{\bf R}_{i})\}$, as follows: the bare atomic energy is
$\epsilon_{a}\equiv \langle  w_{i}\vert  H_{1}\vert w_{i}\rangle$,
the hopping integral $t_{ij}\equiv \langle  w_{i}\vert  H_{1} \vert w_{j}\rangle$, intraatomic-interaction magnitude
$U\equiv \langle  w_{i} w_{i}\vert V_{12} \vert w_{i} w_{i}\rangle$, and
interatomic-interaction magnitude
$K_{ij}\equiv \langle w_{i} w_{j}\vert  V_{12} \vert \\w_{i} w_{j}\rangle$,
where $H_{1}$ is the Hamiltonian for a single particle in the system, and $V_{12}$
represents interparticle interaction. The Wannier functions are expressed
in terms of adjustable Slater atomic  functions, i.e.
$w_{i}({\bf r})=\beta \Psi_{i}({\bf r})-\gamma \sum_{j=1}^{z} \Psi_{j}({\bf r})$
where $z$ is the number of nearest neighbors, $\beta$ and $\gamma$ are mixing coefficients,
and $\Psi_{i}({\bf r})\equiv (\alpha^{3}/\pi )^{1/2}\ \\exp (-\alpha\vert {\bf r}-{\bf R}_{i}\vert)$
is the $1s$ Slater function centered on the site $i\equiv {\bf R}_{i}$.
In the concrete calculations, they are represented by adjustable Gaussians
(e.g. STO-7G basis in dimensions $D=1,2,\mbox{ and }3$, cf. Appendix B). The properties
of those functions have been discussed in detail before. \cite{SpalekPodsiadly,KurzykWojcik}
Here we concentrate on the novel aspects of the approach only. This means we explicitly determine the critical interatomic distance $R=R_c$ for the metal-insulator transition, as well as determine the wave function characteristic as a function  of $R$, both above and below $R_c$. Based on these results we determine scaling properties of physical quantities vs. the parameter $(R-R_c)/R_c$. In effect, we show that not only the wave-functions determination and the electronic correlations are interrelated, but also the former characteristic exhibits a singular behavior at the onset of Mott-Hubbard localization. This last feature of our results represents the principal message of this paper.

\section*{3. Quantum critical behavior of the wave function}

\subsection*{3.1 Overall features of the wave function: evolution of Wannier functions into atomic wave function with $R\rightarrow\infty$}

As said above, the fundamental principle behind our approach is that the wave functions
$\{ w_{i}({\bf r})\}$ determined variationally in the Hilbert space are treated on the same footing as the diagonalization of (\ref{eq:w1})
in the Fock space.
Such diagonalization is possible in an exact manner for nanoscopic and
infinite Hubbard chain $(D=1)$ systems only. \cite{KurzykSpalek,LiebWu} However, also the Gutzwiller wave-function
(GWF) and the Gutzwiller-ansatz (GA) approximations lead to close results in the latter case (see below)
provided the wave functions $\{ w_{i}({\bf r})\}$ are properly readjusted in the correlated
state to achieve the ground-state energy as a global minimum for given lattice parameter $R$.
In other words, the energy of the correlated state is readjusted
iteratively multiple times by adjusting the component atomic-wave-function size $\alpha^{-1}$ and in effect,
also the microscopic parameters $\epsilon_{eff}$, $t_{ij}$, $U$, and $K_{ij}$ determining the many-particle ground state.
As a result of such
iterative approach, we obtain renormalized wave functions $\{w_{i}({\bf r})\}$
and the microscopic parameters, as well as the ground-state energy, and most importantly, an explicit scaling of the physical
properties, all as a function of $R$. Below we analyze first the results obtained for GA for three-dimensional
cubic lattices, before testing the method feasibility for $D=1$ and $2$ situations. The main emphasis is lead
on the novel singular scaling properties near the Mott-Hubbard critical spacing $R=R_{c}$.

\begin{figure}%[H]
%{\bf Figure Captions}
\resizebox{0.55\columnwidth}{!}{%
  \includegraphics{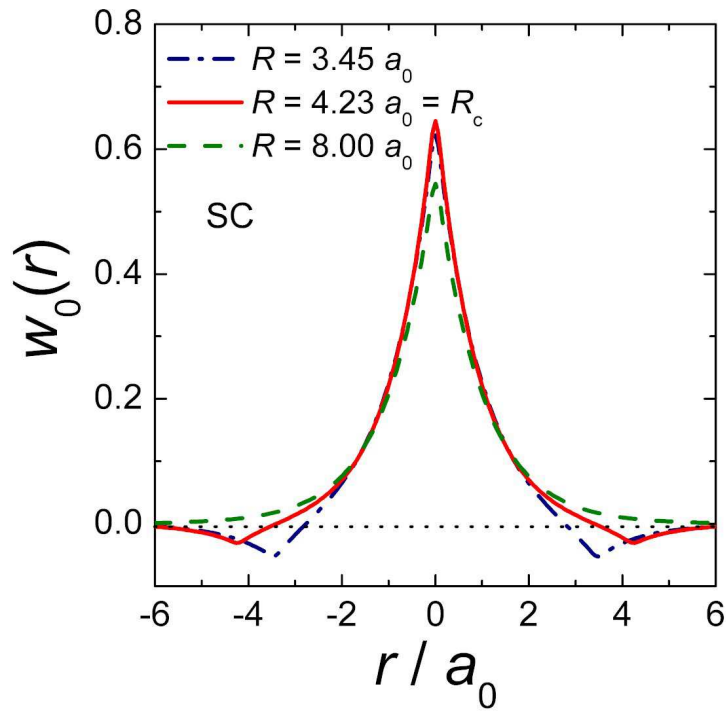}
}
\caption{(Color online) Exemplary shapes of the Wannier function $w_0(\textbf{r})$centered at the
site ${\bf R}_{i}=0$ for simple cubic (SC) lattice along [1,0,0] direction and for three lattice spacings marked:
$R<R_{c}$ (dot-dashed line), $R=R_{c}$ (solid line), and $R>R_{c}$ (dashed line). $a_{0}$
is the $1s$ Bohr radius.}
\label{fig:wannierSC}
\end{figure}

\begin{figure}%[H]
\resizebox{0.55\columnwidth}{!}{%
  \includegraphics{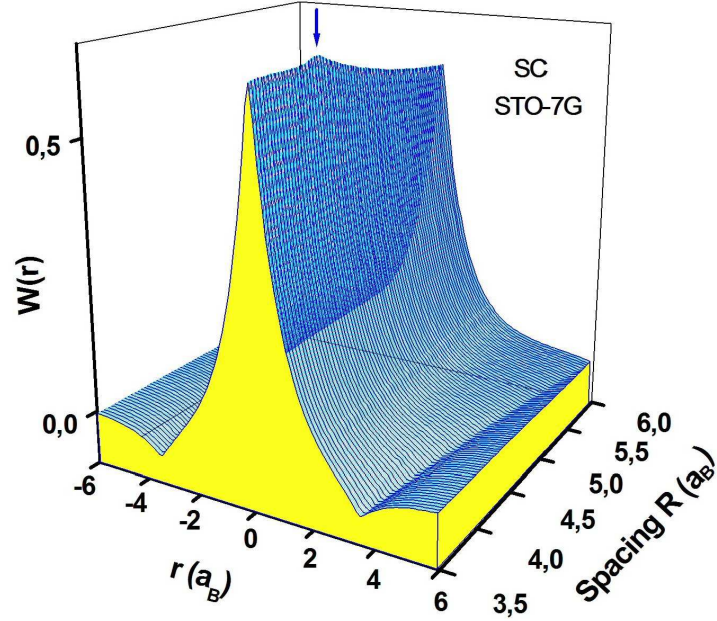}% Here is how to import EPS art
}
\caption{\label{fig:ProfilesSC} Overall space profiles of the renormalized Wannier function for simple-cubic
lattice, as a function of lattice parameter $R$. The arrows mark the position of the singular (critical)
behavior elaborated below in main text (cf. also Fig.~\ref{fig:RelInvSize}. The Wannier function evolves with $R\rightarrow\infty$ ($R\gg R_c$) into atomic 1s wave function.}
\end{figure}

In Fig.~\ref{fig:wannierSC} we plot exemplary Wannier function centered at ${\bf R}_{i}=0$ for simple cubic
(SC) lattice and drawn along [1,0,0] direction and for $R<R_{c}$ (dot-dashed line) $R=R_{c}$ (solid line), and $R\gg R_{c}$
(dashed line). One observes immediately a surprising feature, namely the wave function for $R>R_{c}$ is more extended
than that for $R=R_{c}$ and with the increasing $R$ the size $a=1/\alpha$ gradually reaches its Bohr atomic 1s-state value $a_{0}$.
An overall wave-function profile evolution as a function of $R$ is displayed in Fig.~\ref{fig:ProfilesSC}. On the whole, the wave function seems
to evolve continuously for $R>R_c$ and reaches its atomic $1s$ shape gradually, except for a local maximum
which appears at the specific lattice constant $R=R_{c}$, as marked by the vertical arrow. Before studying this unusual feature let us
note, that our method enables us to study this wave-function evolution in the whole range
of $R$ (across the metal-insulator boundary) as long as the tight-binding approximation is reliable.
This is so, as we shall see and the interesting scaling laws appear
near $R_{c}$ which falls in the limits of large and small interatomic separation, $R\gg\alpha^{-1}$ and $R<\alpha^{-1}$, respectively. The value $R=R_c$ is determined from the point when the renormalized band energy is completely compensated by the Coulomb interaction part and coincides with the correlation function $d\equiv<n_{i\uparrow}n_{i\downarrow}>=0$ in the Gutzwiller-ansatz approximation for the half-filled-band case.

\subsection*{3.2 Wave-function singular behavior at Mott point and scaling properties in the critical regime $R\sim R_c$}

\begin{figure}%[H]
\resizebox{0.55\columnwidth}{!}{%
  \includegraphics{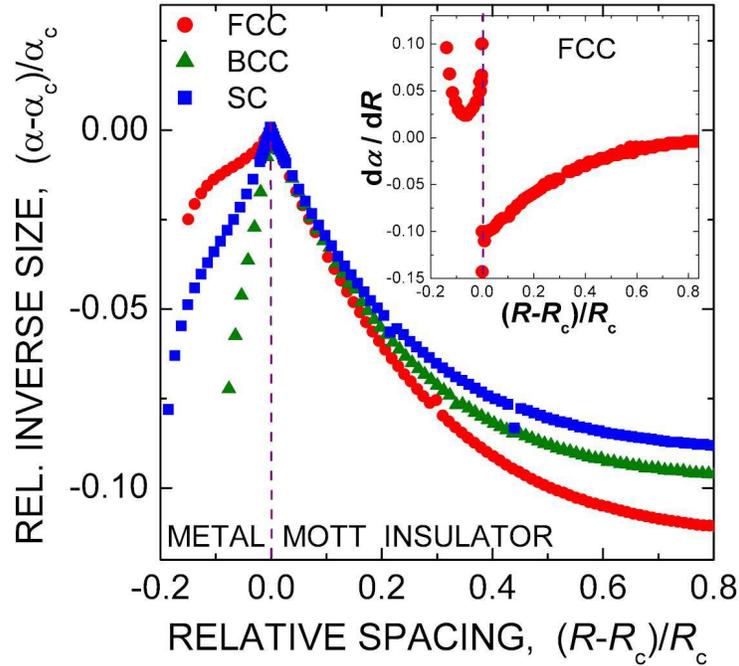}% Here is how to import EPS art
}
\caption{\label{fig:RelInvSize} (Color online) Relative inverse wave function size
$\delta\alpha/\alpha_{C}$ as a function of relative lattice parameter $\delta R/R_{c}$
for simple cubic (SC), body centered cubic (BCC), and face centered cubic(FCC) lattices.
The Mott-Hubbard transition is marked by the vertical dashed line. Inset: derivative
$d\alpha/dR$ vs. $\delta R/R$ with a singularity at $\delta R\equiv R-R_{c}=0$. Note the universal behavior in the range $R\gtrsim R_c$.}
\end{figure}

\begin{figure}%[H]
\resizebox{0.55\columnwidth}{!}{%
  \includegraphics{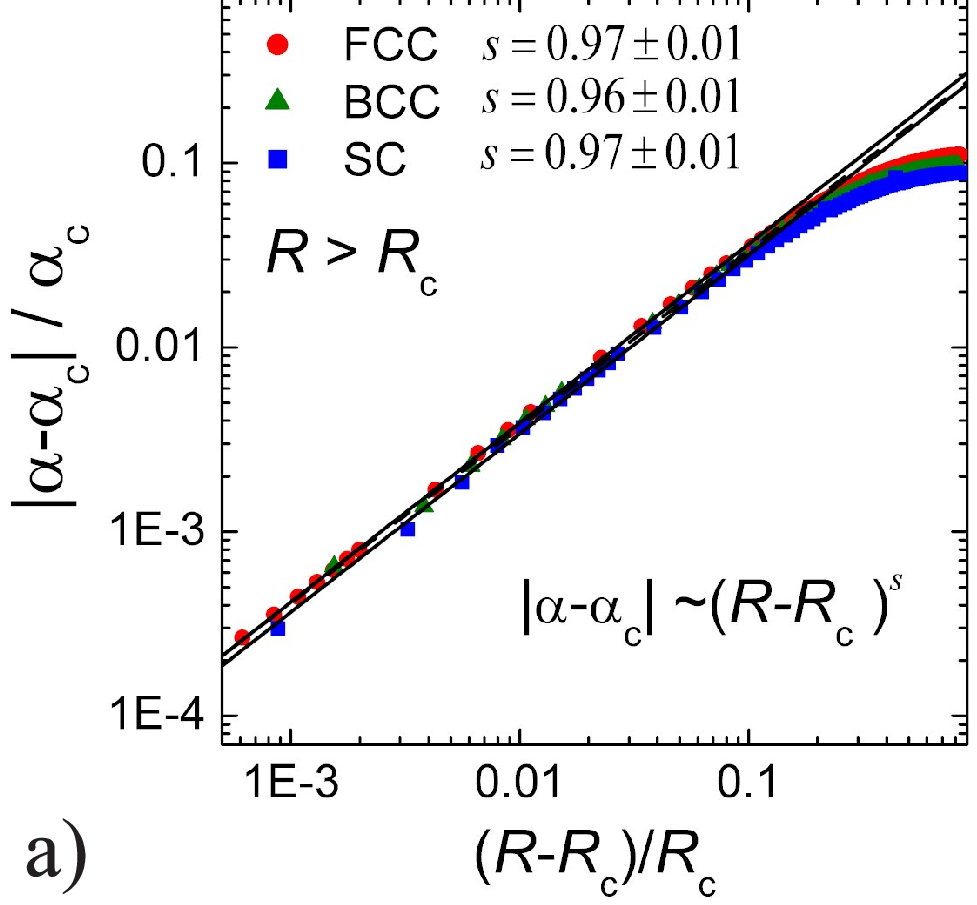}% Here is how to import EPS art
}

\resizebox{0.55\columnwidth}{!}{%
  \includegraphics{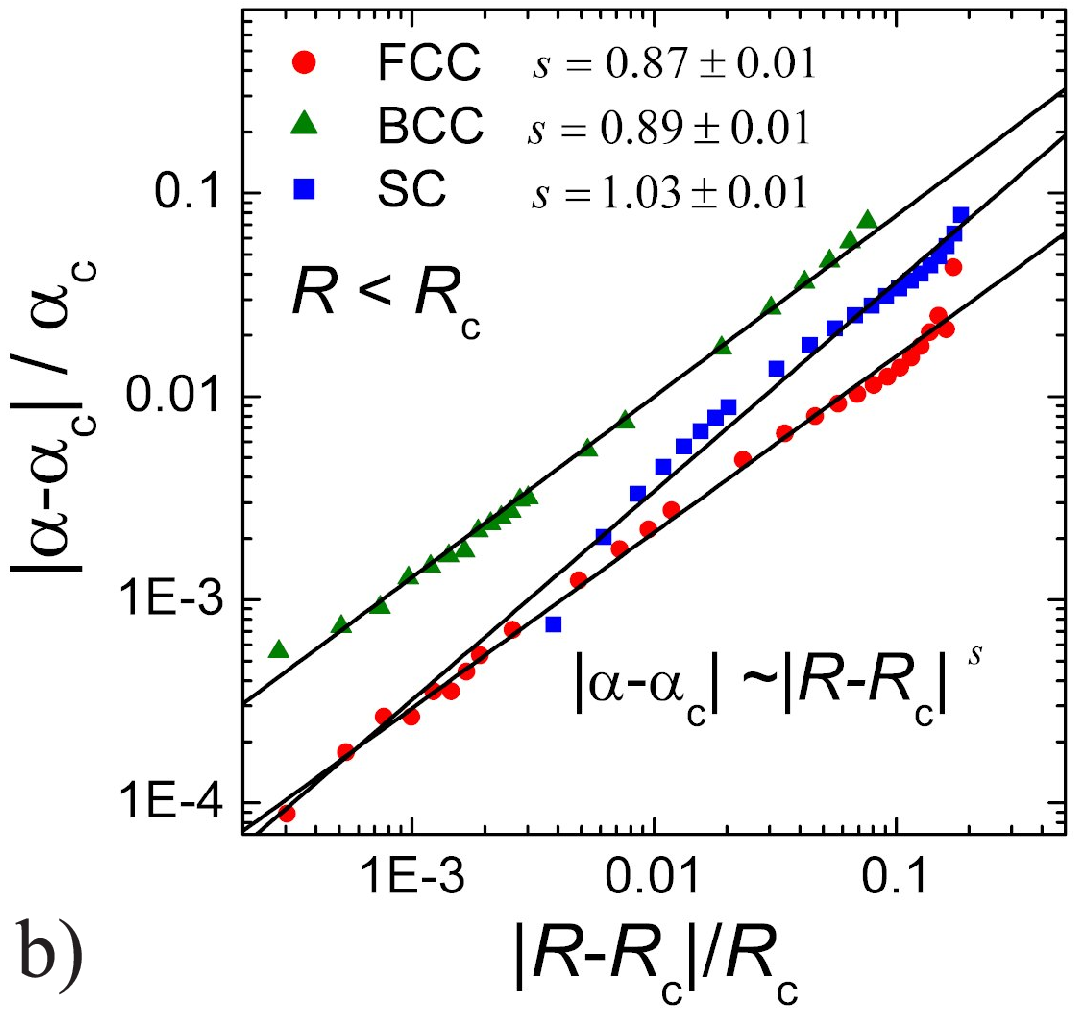}% Here is how to import EPS art
}
\caption{\label{fig:AlfaNormLogScal} (Color online) a) Detailed scaling $\delta \alpha/\alpha$ vs.
$\delta R/R$ and in the doubly logarithmic scale for $R>R_{c}$ (top) and for $R<R_{c}$ (bottom). The straight lines represent the fitted curves $\sim (R/R_{c})^{{\it s}}$. b) The scaling for $R\leq R_c$. Note absence of the degree of universality in the alter (b) case.}
\end{figure}

To study the noted above nontrivial behavior of the wave function in detail, we plot in Fig.~\ref{fig:RelInvSize},
the relative inverse size
$\delta \alpha/\alpha\equiv \vert\alpha(R)-\alpha(R_{c})\vert /\alpha(R_{c})$
of the component atomic functions as a function
of relative lattice spacing $\delta R/R\equiv (R-R_{c})/R_{c}$ in the regime $R\sim R_{c}$. A clear
$d\alpha/dR$ discontinuity (cf. inset) is observed at $R=R_{c}$ for all cubic lattices, a rather unique
and unexpected feature, which is completely absent for the case with bare (unrenormalized) wave functions,
as then $\alpha^{-1}(R)$ would be simply a constant independent of $R$.
Note also that the wave function is the narrowest at the Mott-Hubbard transition, at which $d=<n_{i\uparrow}n_{i\downarrow}>\approx 0$
($d>0$ in the metallic phase). To determine the universality of the behavior, we have replotted
in Fig.~\ref{fig:AlfaNormLogScal}a the result of Fig.~\ref{fig:RelInvSize} for $R>R_{c}$ in a doubly logarithmic scale. One observes a clear
power law scaling $\delta\alpha/\alpha _c\sim [(R-R_{c})/R_{c}]^{{\it s}}$, with ${\it s}= 0.96\pm 0.01$
for not too large $R$ above $R_{c}$. The results for $R<R_{c}$ exhibit also a similar trend with the exponent ${\it s}$ with an overall exponent ${\it s}$ slightly lower $({\it s}\simeq 0.93)$ for $SC$ and $BCC$ lattices i.e. weakly dependent on the type of
cubic lattice selected, as shown in Fig.~\ref{fig:AlfaNormLogScal}b.
This nonuniversal character (cf. also the part for $R<R_{c}$ in Fig.~\ref{fig:RelInvSize}) may be caused by the
circumstance that the tight-binding approximation for the wave function should be extended
to more distant neighbors, as the presence of carriers screens out effectively
the Coulomb interaction (the value $d=\langle n_{i\uparrow}n_{i\downarrow}\rangle$
increases as $R$ decreases). Nonetheless, in spite the lack of simplicity
of the precise scaling for $R<R_{c}$, it is reassuring that the critical-exponent
values for $R<R_{c}$ are close to that for $R>R_{c}$.
This power-law scaling describes the singular behavior of the {\em atomic\/}-wave-function size
$a= 1/\alpha$. Also, the function in Fig.~\ref{fig:RelInvSize} with $R\rightarrow\infty$ does not
reach zero, since $\alpha(R\rightarrow\infty)=a_{0}^{-1}>\alpha(R=R_{c})$.

\begin{figure}%[H]
\resizebox{0.55\columnwidth}{!}{%
  \includegraphics{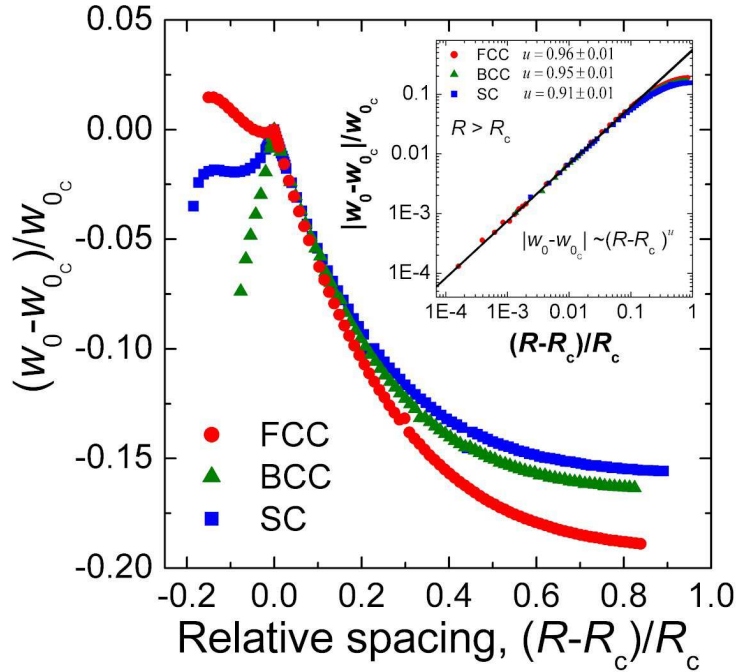}% Here is how to import EPS art
}
\caption{\label{fig:MaxW0} (Color online) Maximum $w_{0}(0,0,0)$
of the Wannier function vs. $\delta R/R_c$. Inset: detailed scaling of the renormalized Wannier function maximum in the doubly logarithmic scale (of same type as in Fig.~\ref{fig:AlfaNormLogScal}). Fitted straight line (for $R>R_{c}$)
represents the function $\sim [(R-R_{c})/R_{c}]^{0.92}$. Note the essential similarity of the scaling above and that shown in Fig.\ref{fig:RelInvSize} and \ref{fig:AlfaNormLogScal}.}
\end{figure}

One can also study the related properties of the renormalized Wannier function directly.
For that purpose we have displayed in Fig.~\ref{fig:MaxW0} the maximal value of $w_{i}({\bf{r}=0})$
as a function of the scaling parameter $(R-R_{c})/R_{c}$. Again, also for that quantity,
we observe a similar type
of scaling as for $\alpha^{-1}$, with the critical exponent $\alpha =0.92\pm 0.01$ (cf.
inset). However, the corresponding scaling law in the regime $R<R_{c}$ is again not as clear,
as that for $R>R_{c}$ (cf. Figs. ~\ref{fig:RelInvSize} and ~\ref{fig:MaxW0}). Note that close to $R=R_c$ we have $(\alpha - \alpha _c)/\alpha _c\simeq (a-a_c)/a_c$, where $a$ is the renormalized Bohr radius in the correlated state. In other words both $\delta\alpha/ \alpha _c$ and $\delta \alpha _c$ scale the same way with $\delta R/R_c$.

\begin{figure}%[H]
\resizebox{0.55\columnwidth}{!}{%
  \includegraphics{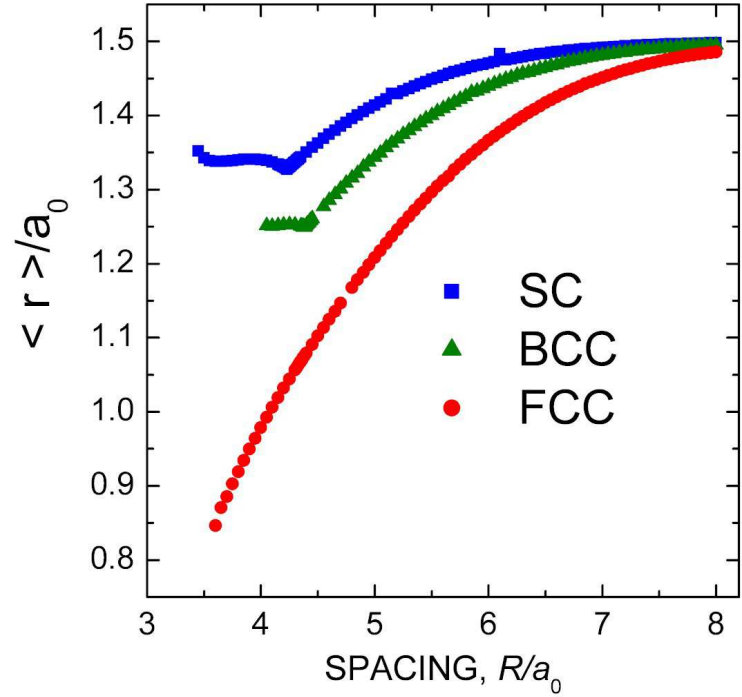}% Here is how to import EPS art
}
\caption{\label{fig:AverageR} (Color online) First moment (an average size) of the renormalized Wannier function for cubic lattices. Note the same type of discontinuity as in Fig.~\ref{fig:RelInvSize} for SC and BCC lattices.}
\end{figure}

\begin{figure}%[H]
\resizebox{0.55\columnwidth}{!}{%
  \includegraphics{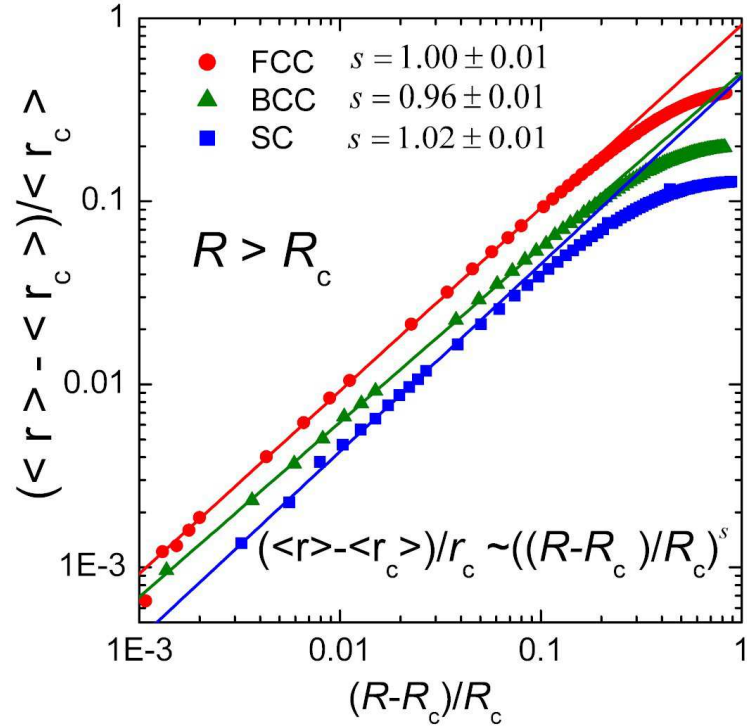}% Here is how to import EPS art
}
\caption{\label{fig:RelAverSize} (Color online) Relative average size $(\left\langle r\right\rangle-\left\langle r_c\right\rangle)/\left\langle r_c\right\rangle$ of the renormalized Wannier function for $SC$, $BCC$, and $FCC$ cubic lattices as a function
of relative lattice constant $\delta R/R_c$. The discontinuities appear at the critical value
$R_{c}$. The $R\rightarrow\infty$ asymptotic value of $\langle r\rangle=1.5 a_{0}$
is that for atomic $1s$ wave function, with $a_{0}$ being the Bohr radius.}
\end{figure}

There is no obvious quantity like $\alpha^{-1}$ characterizing the width of
the renormalized Wannier function $w_{i}({\bf r})$. This is the reason why we calculated an overall (average) size of the renormalized Wannier function defined as $\langle r\rangle\equiv \int r w^2({\bf r}) d^{3}{\bf r}$,
and have plotted it in Fig.~\ref{fig:AverageR} as a function of $\delta R/R_{c}$ for the three cubic lattices.
While we observe anticusp behavior of $\langle r\rangle$ at $R=R_{c}$ for $SC$ and $BCC$ lattices,
there is no clear sign of that happening for the $FCC$ case. However, one has to realize,
that calculation of $\langle r \rangle$ involves
a three-dimensional integration over space in the situation when the Wannier
functions are strongly anisotropic. \cite{KurzykSpalek} Because of this specific reason, we regard the scaling
of $w_0(\textbf{r})$ maximum shown in Fig.~\ref{fig:MaxW0} as more direct, as since involves the wave function characteristic
which is not averaged out and, obviously, independent of the spatial direction.
Nonetheless, in Fig.~\ref{fig:RelAverSize} we display the corresponding scaling of the relative difference $(\left\langle \bf{r}\right\rangle-\left\langle \bf{r}_{c}\right\rangle)/r_c$ for $R>R_c$. The less systematic behavior even for $R>R_c$ does not allow us to draw any definite conclusions for that regime. Note that for $R>R_c$ even in the case of FCC lattice, where $\left\langle \bf r\right\rangle$ is approximately continuous when crossing $R_{c}$, we observe the same type of scaling, as that discussed for SC and BCC lattices. This means again that the scaling in the Mott insulating phase is induced by the intersite Coulomb interaction. 

\subsection*{3.3 Scaling property of ground-state energy and physical meaning of the scaling}

\begin{figure}%[H]
\resizebox{0.55\columnwidth}{!}{%
  \includegraphics{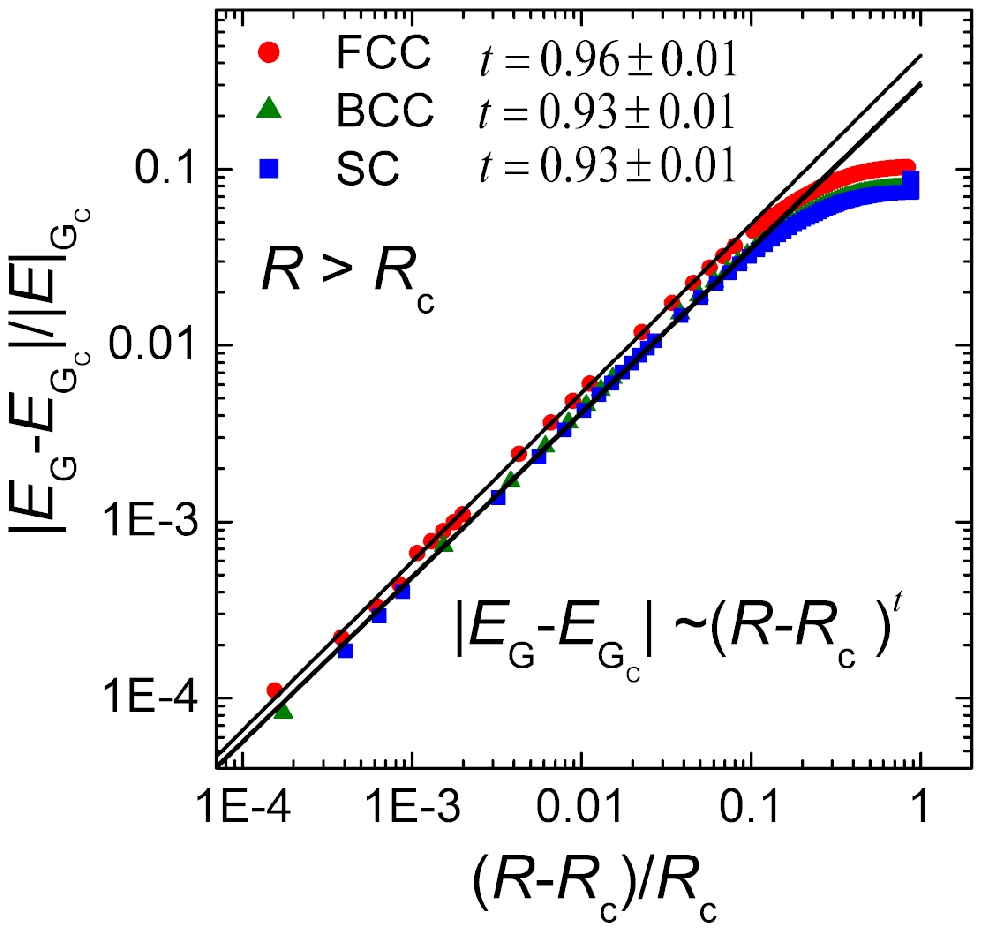}% Here is how to import EPS art
}
\caption{\label{fig:EnergyScal} (Color online) Same scaling as in Fig.~\ref{fig:AlfaNormLogScal} but for the ground state
energy; the fitted exponents are listed for the three cubic lattices. In the asymptotic limit $\delta R/R_c>>1$ the three values of relative energy merge, since the system approaches the atomic limit. Close to the atomic limit the scaling is not obeyed. This means that a collective Mott state sets in with $R\rightarrow R_c^{0+}$.}
\end{figure}

The scaling with varying $\delta R/R$ is also observed for the ground
state energy, as shown in Fig.~\ref{fig:EnergyScal}. The approximate dependence
$\sim (\delta R/R)^{t}$ of $E_{G}$ when approaching
the atomic limit can be attributed to the dominant role of the Coulomb repulsion (note the absolute value of the energy).

\begin{figure}%[H]
\resizebox{0.55\columnwidth}{!}{%
  \includegraphics{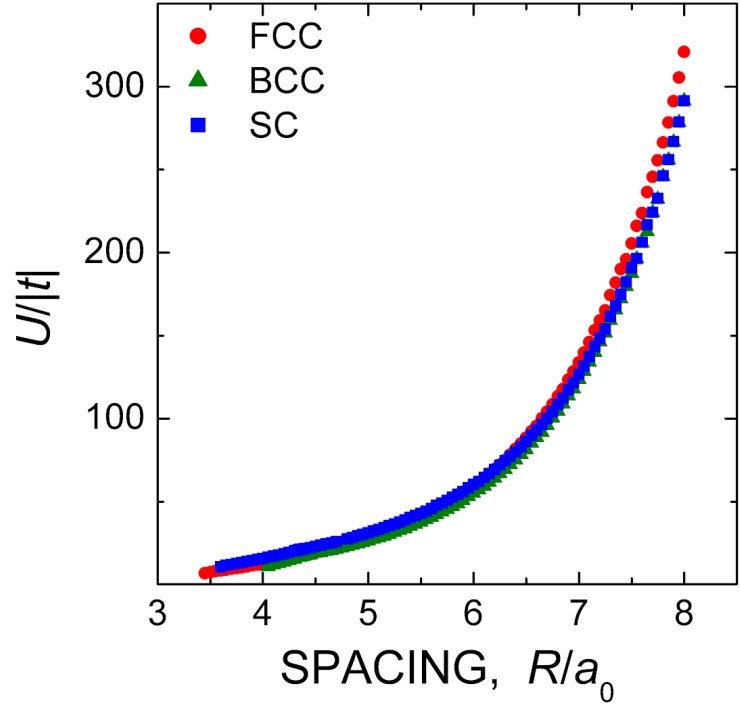}% Here is how to import EPS art
}
\caption{\label{fig:U_t3D} (Color online) The $U/|t|$ ratio as a function of the interatomic spacing for cubic lattices showing its monotonic increase upon increasing $R$, unlike the scaling of the Wannier function characteristics.}
\end{figure}

To summarize this Section, both the Figs ~\ref{fig:AlfaNormLogScal} and~\ref{fig:RelAverSize} demonstrate (for $R\geq R_{c}$), respectively the appearance of a overall scaling of the $R^{\pm n}$ type,
with $n\sim 1$.
This dependence could be seen in the pressure dependence of the orbital size when close to the critical value
$R=R_{c}$. However, the relative changes of $\alpha$ and $\left\langle r\right\rangle$ are rather subtle and the question remains 
if they can become observable in the present day e.g. scanning tunneling observations of the localized
orbitals in the Mott insulating state. For such purpose, relevant are the plots of the maximal value $w_{0}(0)$
of the Wannier function relative to that at $R=R_{c}$ (cf. Fig.~\ref{fig:MaxW0}). The particle occupancy change
at $r=0$ reaches up to
$20\%$ of the peak value upon change of $20-25\%$ of the lattice constant either way, so it could become observable.
On the basis of our results one can also say, that our approach
provides evidence for effects in the Wannier function, which are pronounced
already in the Mott-Hubbard insulating state. In other words, the Hubbard split subband
picture of the Mott insulator is concomitant with a strong renormalization of the wave
function characteristics when approaching the transition point. This critical behavior does
not alter the monotonic increase of the $U/|t|$ ratio with the increasing lattice constant \cite{KurzykWojcik}, as shown explicitly in the Fig.~\ref{fig:U_t3D} for all three cubic lattices. The almost exponential growth of this ratio for $R>4a_0$ is a signature of entering into the strong-correlation limit. Also, the presence of the scaling demonstrates the onset of the collective character of the quantum critical regime, in which the Mott insulating state cannot be regarded as a collection of atomic states with unpaired spins, at least when $R\sim R_c$.

\section*{4. Effect of lattice dimensionality: feasibility of the method and the Gutzwiller-ansatz approximation vs. the original Mott criterion}

\begin{figure}%[H]
\resizebox{0.55\columnwidth}{!}{%
  \includegraphics{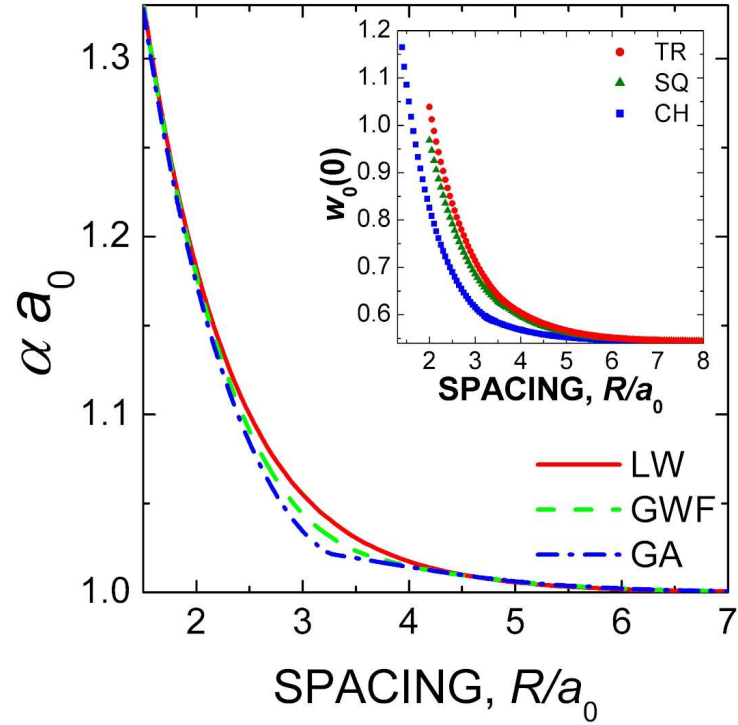}% Here is how to import EPS art
}
\caption{\label{fig:InvSizeCH} (Color online) Inverse size of the renormalized atomic wave function
for linear chain (CH) vs. $R/a_{0}$ within the exact Lieb-Wu (LW), the Gutzwiller-wave-function
(GWF), and the Gutzwiller-ansatz (GA) solutions. No singular behavior is observed at any $R$.
Inset: Wannier-function maximum $w_{0}(0)$ vs. $R/a_{0}$ for linear chain (CH),
square (SQ), and triangular lattices (TR), with no discontinuity detected in either of the cases. The scaling in GA approximation discussed above does not apply here.}
\end{figure}

Finally, we examine the role of lattice dimensionality on the singular behavior of the
wave function. For that purpose, we plot in Fig.~\ref{fig:InvSizeCH} the inverse wave-function size
$\alpha^{-1}$ vs. $R/a_{0}$
for the linear chain within the modified exact Lieb-Wu (LW), \cite{LiebWu} Gutzwiller-
wave-function (GWF), \cite{Metzner} and Gutzwiller-ansatz (GA) \cite{KurzykWojcik,KurzykSpalek,Gutzwiller} solutions, respectively.
None of the three methods provides in $D=1$ case (and GA also in the $D=2$ case) any type of singular
behavior observed in Figs.~\ref{fig:RelInvSize} and ~\ref{fig:MaxW0} for the $D=3$ case.
What is more important, all the three methods provide quite similar results for $D=1$, i.e.
there is no critical behavior of the type discussed above. \cite{LiebWu} This circumstance
provides us with some confidence, that modified by us GA method (including the interatomic interaction exactly in the Mott insulating state) bears some
physical relevance to the MIT transition treated in mean-field approximation.
MIT transition for $D=1$ and $D=2$ lattices which appears in the standard (mean-field) Gutzwiller ansatz,
does not show up in the behavior of the renormalized wave function in real space. This is one of the main supplementary results of this paper.
To illustrate that, we plot in Fig.~\ref{fig:InvSizeCH} 
exemplary Wannier function vs. $R/a_{0}$ for
triangular (TR) and square (SQ) lattices. The results for $D=2$ are obtained for GA solution.
Again, no critical behavior discussed above is observed. 

\begin{figure}%[H]
\resizebox{0.55\columnwidth}{!}{%
  \includegraphics{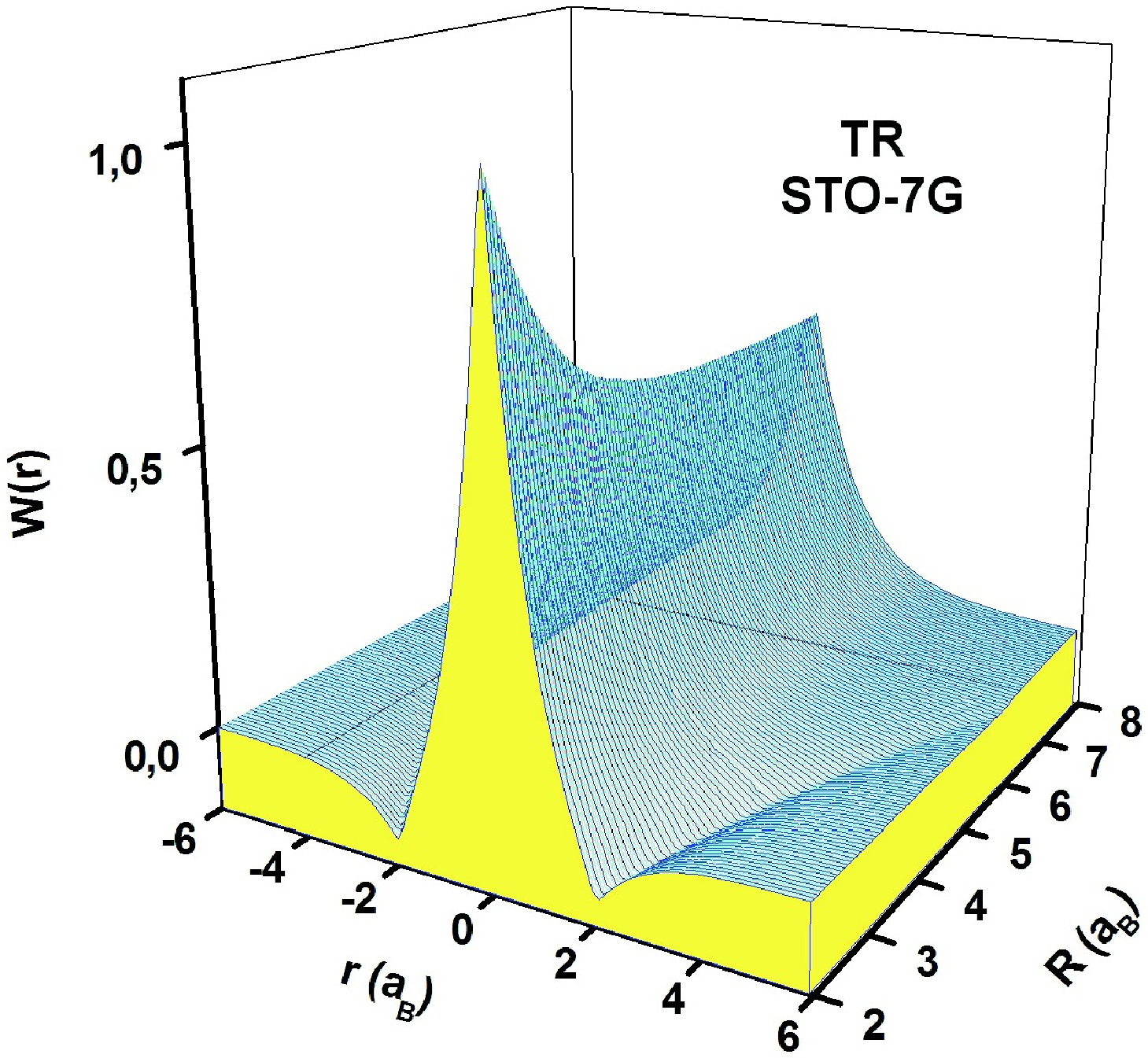}% Here is how to import EPS art 
}

\resizebox{0.55\columnwidth}{!}{%
  \includegraphics{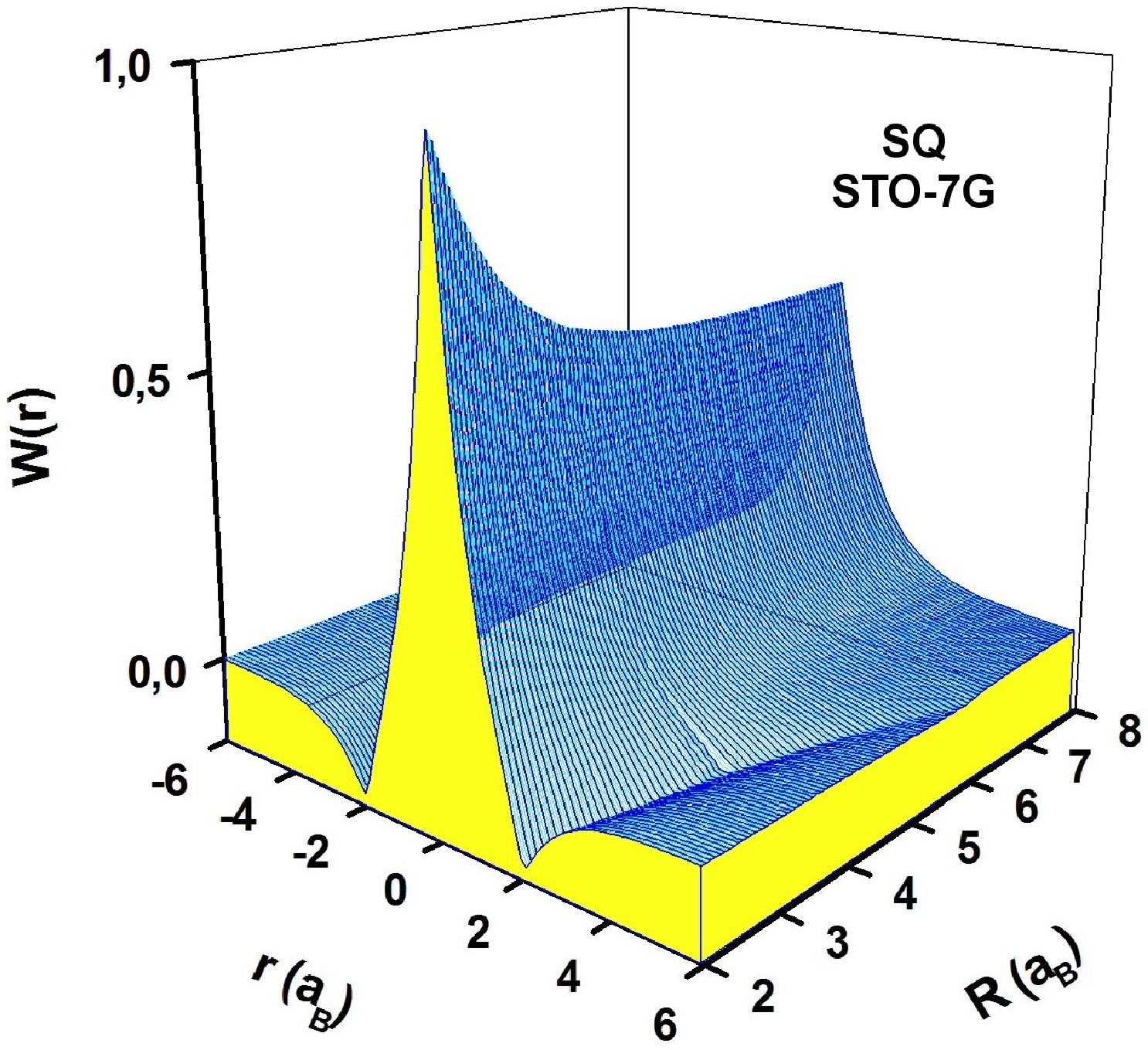}% Here is how to import EPS art 
}
\caption{\label{fig:RenWannSQTR} (Color online) Space profiles of the Wannier functions for triangular (TR, top) and square (SQ, bottom) planar lattices, both as a function of lattice constant $R$. No singular behavior within GA approach is observed. Note a gradual evolution with the increasing of $R$ of the Wannier functions into the corresponding atomic (1s) form.}
\end{figure}

On the basis of these results we draw the conclusion,
that within our approach the critical behavior of the Wannier function appears only for high-dimension ($D>2$) lattices. This circumstance is in accord with the notion that GA results are more realistic in high dimensions.
The striking coincidence of these results with the critical dimensionality $(D>2)$ for the
onset of Landau-Fermi liquid stability should therefore be mentioned. \cite{Houghton}
The metallic $3D$ state of the present approach is represented
by an {\em almost localized Fermi liquid\/} \cite{SpalekDatta} (ALFL), which differs from the Landau Fermi liquid by a strong (and spin dependent) mass enhancement factor, as well as by a presence of metamagnetism. \cite{SpalekWojcik}

\begin{figure}%[H]
\resizebox{0.55\columnwidth}{!}{%
  \includegraphics{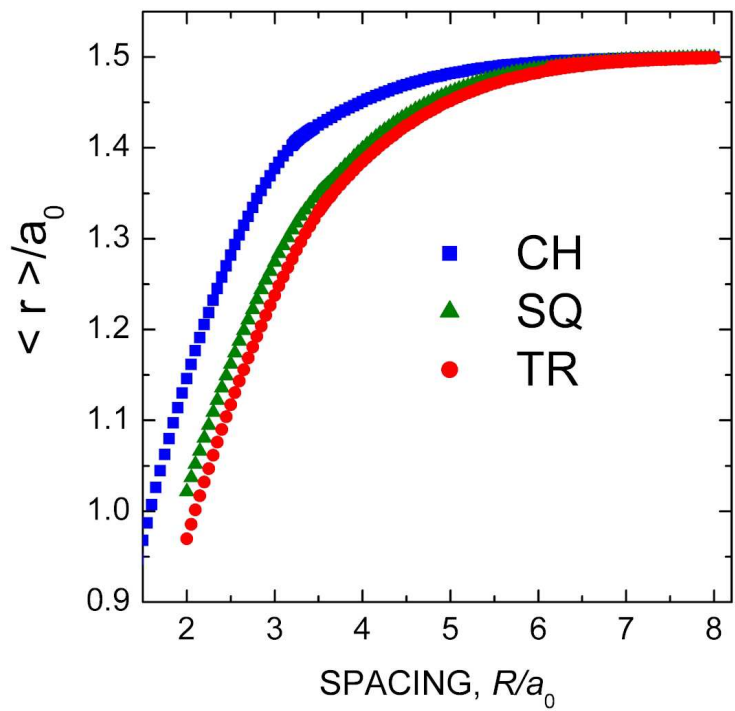}% Here is how to import EPS art
}
\caption{\label{fig:AverSizeWannCH} (Color online) Average size of the renormalized Wannier function
for one- (CH), and two-dimensional (SQ, TR) lattices. No singular behavior observed apart from a slight slope change for the linear chain.}
\end{figure}

As a supplementary information, we provide in Fig.~\ref{fig:AverSizeWannCH} the average size $\langle r \rangle$
of the renormalized Wannier function for linear chain (CH) and the two-dimensional lattices, SQ and TR.
As is the case in Fig.~\ref{fig:RenWannSQTR}, no singular character has been observed for this quantity.
These results confirm again that our results
are sensitive to the lattice dimensionality and nontrivial properties appear only in the high-dimension
limit, where the mean-field (Gutzwiller) approximation may be regarded as at least
qualitatively correct. \cite{SpalekWojcik} Here they appear already for $D=3$.

One additional basic feature of our approach should be mentioned. Namely, in Table I
we list characteristics of the Mott-Hubbard transition for the cubic lattices.
These results intercorrelate nicely
with the Mott criterion in the following manner. The critical carrier concentration for the Mott transition in the present situations is $n_{C}=1/R_{c}^{3}$
for SC, $2/R_{c}^{3}$ for BCC, and $4/R_{c}^{3}$ for FCC. Thus $n_{C}=n/R_{c}^{3}$,
with $n=1,2,\mbox{ and }4$, respectively. The Mott criterion discussed in qualitative terms in Appendix A, takes for those lattices the form:
$n_{C}^{1/3}\cdot a= n^{1/3}/(R_{c}\alpha_{C})=0.20$, $0.26$, and $0.33$,
for SC, BCC, and FCC lattices, very close values to that obtained originally by Mott, \cite{Mott} and rederived in an elementary manner in Appendix A.
This connection provides an additional argument for a qualitative correctness of our approach
in the sense, that a well defined value of $R_{c}$ obtained from GA solution,
corresponds nicely with the historic criterion introduces by Mott. \cite{Mott} Parenthetically, the agreement (cf. Table 1) relates nicely the Mott and Hubbard criteria for MIT.

\begin{table}%[H]
\begin{center}
\caption{Microscopic parameters near the mean-field critical point calculated for the cubic lattices.}
\begin{tabular}{c|c|c|c}
\hline\hline
Struct. & $\alpha a_{0}$ & $R_{c}/a_{0}$ & $(U/W)_{C}$ \\
\hline
SC & 1.099 & 4.236 & 1.337 \\

BCC & 1.109 & 4.384 & 1.080 \\
FCC & 1.128 & 4.351 & 0.880
\end{tabular}
\end{center}
\end{table}

\section*{5 Outlook}
\subsection*{5.1 A brief overview of the method combing first-  and second-quantization schemes}

Our whole method starts with the definition of the field operators

\begin{center}
$
\left\{ \begin{array}{l}
 \mathord{\buildrel{\lower3pt\hbox{$\scriptscriptstyle\frown$}} 
\over \Psi } ({\bf{r}}) = \sum\limits_i {\Phi _i ({\bf{r}})(a_{i \uparrow } ,a_{i \downarrow } ),}  \\ 
 \mathord{\buildrel{\lower3pt\hbox{$\scriptscriptstyle\frown$}} 
\over \Psi } ^\dag  ({\bf{r}}) = \sum\limits_i {\Phi _i ({\bf{r}})\left( \begin{array}{l}
 a_{i \uparrow }^\dag   \\ 
 a_{i \downarrow }^\dag   \\ 
 \end{array} \right),}  \\ 
 \end{array} \right.
$
\end{center}

and of the corresponding Hamiltonian (1) in the second quantization representation. The method of approach is equivalent to that starting from N-particle Schr\"odinger equation provided the single-particle basis $\left\{ \Phi_i(\bf{r})\right\}$ is complete in the quantum-mechanical sense. The basis $\left\{ \Phi_i(\bf{r})\right\}$ can be arbitrary, even not orthonormal \cite{Feynman}, although the choice of orthogonal basis simplifies the whole procedure. Within this scheme, the starting Hamiltonian (1) represents \textit{a model}, as it includes only a single 1s Wannier function $\left\{w_i(\textbf{r})\right\}$ per atomic site (no other orbitals included)..

The question is how to select the correct wave-function basis $w_i(\textbf{r})$ for this model. This question has been addressed a number of times in a general case at the original stage of setting the quantum theory. \cite{Fock} In general, by introducing the occupation-number representation we determine combinatorially (or algebraically) the number of single-particle configurations and the resulting interaction energy as an average over the set of these coherent configurations (a multiconfigurational average). The single-particle wave function on other hand, should be adopted variationally in the resultant multiparticle state, particularly if the interaction energy is comparable to the single-particle (band) part, as no perturbational approach starting solely from the single-particle part is appropriate. Conversely, the single-particle wave-function size readjusted in the correlated state represents an additional factor lowering the system energy. \cite{Schrodinger} This procedure has been carried out successfully first in the case of the He atom and H$_{2}$ molecule. \cite{Bethe} Minimally, determining the system evolution as a function of model parameters, which is insufficient from the physical point of view. Obviously, such a variational wave-function readjustment is carried out already in e.g. LDA method. \cite{Hohenberg} Our method of approach demonstrates explicitly, albeit in a model situation, that the single-particle wave function variational readjustment only is insufficient, as an intrinsically different ingredient, the two-particle correlation function, is ignored in the process. The LDA and related methods are certainly valid (and often sufficient) when the two-particle correlations can be approximated by a product containing density of particles only, i.e. in the weakly or moderately regime of correlated particles.

\subsection*{5.2 Conclusions}

In summary, we have extended the standard treatment of the Mott-Hubbard transition
by incorporating into
it a self-consistent scheme the single-particle wave-function renormalization, as well as demonstrating its
singular behavior for three-dimensional lattices within the Gutzwiller approximation.
These effects are absent in lower dimensions ($D<3$).
We have also related the critical interatomic distance $R_c$ to the original Mott criterion. The correlation effects in the single-particle wave-function-shape
develop already in the Mott-Hubbard insulating phase.
The wave-function characteristic length $\alpha^{-1}$ (or its average size $\langle r \rangle$)
play the role of a coherence length for a single-fermion placed in the
mean field of all other particles. In connection with this one has to mention the works on the scaling theory of Anderson transition, where the coherence length scales with the magnitude of interaction with the exponent $\nu =-1$. In our case roughly $(\left\langle r\right\rangle - \left\langle r_c\right\rangle)\sim (R-R_c)\sim (K_c-K)/K_c^2$ for $R>R_c$, where $K_c$ is the value of $K$ for $R=R_c$. This means the scaling is different in the present case and differentiates between the Mott and Anderson transitions.

The present approach can also be extended to the orbitally degenerate $3d$ states along the lines
proposed earlier. \cite{Klejnberg,Bunemann}
Important is the circumstance that the present method, although still at a simple modeling stage, avoids the notoriety of counting twice the
interparticle interaction, present in any of the current ab initio approach, and goes beyond the parameterized-model calculations \cite{KurzykWojcik,KurzykSpalek}, as it provides the system evolution as a function of lattice constant. Also, it is of fundamental importance to try to extend the present {\em hypothesis of a quantum critical scaling\/} to check
if it holds beyond a mean-field (high-dimensional) treatment
of strongly correlated fermions near the Mott-Hubbard transitions.

\subsection*{Acknowledgments}

The authors are grateful to Krzysztof Byczuk for interesting points on DMFT method.
The work was supported by the Grant No. NN 202 128 736 from Ministry of Science
and Higher Education. Also, the research is performed under the auspices of the
ESF Network INTELBIOMAT and the National Network {\em Materials with Strongly Correlated Electrons\/}. The authors acknowledge also discussions with Dr. Adam Rycerz on and adjustable Gaussian basis. We are also grateful to Dr. Dominique Delande from Universite Paris VI for turning our attention to Ref.\cite{Vollhardt}.

\section*{Appendix A: Elementary meaning of the original Mott criterion}

The Mott criterion of electron localization can be visualized in an elementary manner from the following reasoning of the gas instability. The kinetic energy per particle in $3D$ ideal gas is
\begin{equation}
\overline{\epsilon}=(3/5)\epsilon_{F}=\frac{3}{5}\frac{\hbar^{2}}{2m^{*}}
\left( 3\pi^{2}\frac{N}{V}\right)^{2/3}\sim\rho^{2/3}\,,
\end{equation}
where $\epsilon_{F}$ is the gas Fermi energy $\rho=N/V$-the particle
density and $m^{*}$-the effective mass in the gas. The electron-electron
interaction at the onset of localization can be estimated as
\begin{equation}
\epsilon_{e-e}=\frac{1}{2}\frac{e^{2}}{\epsilon d_{e-e}}=
\frac{e^{2}}{2\epsilon}\rho^{1/3}
\end{equation}
where the classical interparticle distance has been taken as
$d_{e-e}=(N/V)^{1/3}=\rho^{-1/3}$.

Now, the gas instability point is defined for a critical density
$\rho=\rho_{c}$ for which $\overline{\epsilon}=\epsilon_{e-e}$.
This condition leads to identity
\begin{equation}
\left( \frac{\hbar^{2}}{m^{*}e^{2}}\epsilon\right) \rho_{c}^{1/3}=
\frac{5}{3}\frac{1}{\left( 3\pi^{2}\right)^{2/3}}
\simeq 0.17\,,
\end{equation}
In other word
\begin{equation}
a_{B}\rho_{c}^{1/2}\sim 0.2
\end{equation}
which represents the Mott-Wigner criterion of localization,
with $a_{B}$ being the effective Bohr radius of $1s$-type
of atomic state for electrons in the medium with dielectric constant
$\epsilon$ and with effective mass $m^{*}$. For $\rho<\rho_{c}$
the Coulomb repulsion dominates $(\overline{\epsilon}<\epsilon_{e-e})$.

\section*{Appendix B: Approximation of Slater 1s functions by an adjustable STO-nG basis}

In our previous paper \cite{KurzykWojcik} regarded as Part I we have not defined explicitly the adjustable Gaussian basis. The details of this are provided below.

The method of approximating the molecular orbitals by Gaussians was introduced by Hehre at al. \cite{Hehre}. In this representation the Gaussian orbitals have been introduced in the form
\begin{equation}
g_i^{(a)} ({\bf{r}}) = \left( {\frac{{2\Gamma _a^2 }}{\pi }} \right)^{3/4} e^{ - \Gamma _a^2 |{\bf{r}} - {\bf{R}}_i |^2 },
\end{equation}
where ${\bf{R}}_i $ is the reference atomic site and the index "a" labels different functions. In the basis STO-nG the 1s wave function is written as a linear combination (contraction) of n such Gaussians. As we would like to have a variable-size orbitals, we make a readjustment $\Gamma _a  \to \Gamma '_a  \equiv \alpha \Gamma _a $. In effect, the atomic 1s wave function has the form
\begin{equation}
\Psi _i ({\bf{r}}) = \alpha ^{3/2} \sum\limits_{a = 1}^n {\beta _a g_i^{(a)} ({\bf{r}})}.
\end{equation}
The parameters $\beta _a$ and $\Gamma '_a$ are determined from minimalization of energy of a single atom described by the following Hamiltonian in the atomic units:
\begin{equation}
H_i \mathop  = \limits^{a.u.}  - \nabla ^2  - \frac{2}{{|{\bf{r}} - {\bf{R}}_i |}},
\end{equation}
and are independent of $\alpha$. In Ref. \cite{Fernandez} one can find the expressions for various integrals involving Gaussian basis. Particularly important for us are the following integrals:

1.	The overlap integral between sites $i$ and $j$
\begin{equation}
S_{ab} (R_{ij} ) \equiv \left\langle {g_i^{(a)} } \right|\left. {g_j^{(b)} } \right\rangle  = \left( {\frac{{2\Gamma _a \Gamma _b }}{{\Gamma _a^2  + \Gamma _b^2 }}} \right)^{3/2} e^{ - \frac{{\Gamma _a^2 \Gamma _b^2 }}{{\Gamma _a^2  + \Gamma _b^2 }}R_{ij} }.
\end{equation}
2. The kinetic energy matrix elements
\begin{equation}
\left\langle {g_i^{(a)} } \right| - \nabla ^2 \left| {g_j^{(a)} } \right\rangle  = \left( {\frac{{\Gamma _a \Gamma _b }}{{\Gamma _a^2  + \Gamma _b^2 }}} \right)^{7/2} 
\end{equation}
$$
\left( {\frac{3}{2}\left( {\Gamma _a^2  + \Gamma _b^2 } \right) - R_{ij}^2 \Gamma _a^2 \Gamma _b^2 } \right)e^{ - \frac{{\Gamma _a^2 \Gamma _b^2 }}{{\Gamma _a^2  + \Gamma _b^2 }}R_{ij} }
$$
3. The attractive Coulomb-interaction integral
\begin{equation}
\left\langle {g_i^{(a)} } \right|\frac{2}{{|{\bf{r}} - {\bf{R}}_k |}}\left| {g_j^{(b)} } \right\rangle  =
\end{equation}  
$$- 2S_{ab} (R_{ij} )\frac{{{\rm{erf}}\left( {\left( {\Gamma _{\rm{a}}^{\rm{2}}  + \Gamma _{\rm{a}}^{\rm{2}} } \right)R_{ikj}^{ab} } \right)}}{{R_{ikj}^{ab} }},
$$
where 
\begin{equation}
{\rm{erf}}(x) = \frac{2}{{\sqrt \pi  }}\int\limits_0^x {e^{ - t^2 } dt}
\end{equation}
is the error function and
\begin{equation}
R_{ikj}^{ab}  \equiv \left| {{\bf{R}}_k  - \left( {{\bf{R}}_i  + \frac{{{\bf{R}}_{ij} }}{{1 + (\Gamma _a /\Gamma _b )^2 }}} \right)} \right|.
\end{equation}
Using the above formulas, the ground state energy of hydrogen atom takes the form
\begin{equation}
\left\langle {\Psi _i } \right|H_i \left| {\Psi _i } \right\rangle  = 
\end{equation}
$$
\sum\limits_{a,b} {\beta _a \beta _b \left( {\frac{{3 \cdot 2^{5/2} \left( {\Gamma _a \Gamma _b } \right)^{7/2} }}{{\left( {\Gamma _a^2  + \Gamma _b^2 } \right)^{5/2} }} - \frac{{2^{7/2} \left( {\Gamma _a \Gamma _b } \right)^{3/2} }}{{\pi ^{1/2} \left( {\Gamma _a^2  + \Gamma _b^2 } \right)}}} \right)}
$$
Minimizing (15) under the condition of proper normalization, i.e.
\begin{equation}
\left\langle {\Psi _i } \right|\left. {\Psi _i } \right\rangle  = \sum\limits_{a,b} {\beta _a \beta _b \left( {\frac{{2\Gamma _a \Gamma _b }}{{\Gamma _a^2  + \Gamma _b^2 }}} \right)^{3/2} }  = 1,
\end{equation}
we obtain the values of the parameters $\beta _a$ and $\Gamma _a $ with $a=1,...,7$. Those parameters for the Slater 1s function are listed in Table II.

\begin{table}%[H]
\begin{center}
\caption{Values of parameters $\beta _a$ and $\Gamma _a$ for $a=1,...,7$ for 1s Slater wave function obtained in (STO-nG) representation for $n=3,5, $ and 7, respectively.}
\begin{tabular}{c|c|c|c|c|c|c}
\hline\hline
 & \multicolumn{2}{c|} {STO-3G} & \multicolumn{2}{c|} {STO-5G} & \multicolumn{2}{c} {STO-7G} \\
\hline
 a & $\beta _a$ & $\Gamma _a$ & $\beta _a$ & $\Gamma _a$ & $\beta _a$ & $\Gamma _a$\\
\hline
 1 & 0.7079 & 0.4037 & 0.4862 & 0.3429 & 0.3348 & 0.3073\\
 2 & 0.3460 & 0.8920 & 0.4687 & 0.6490 & 0.4948 & 0.5342\\
 3 & 0.0692 & 1.9706 & 0.1446 & 1.2283 & 0.2219 & 0.9285\\
 4 &        &        & 0.0307 & 2.3249 & 0.0674 & 1.1638\\
 5 &        &        & 0.0094 & 4.4003 & 0.0188 & 2.8050\\
 6 &        &        &        &        & 0.0039 & 4.8755\\
 7 &        &        &        &        & 0.0018 & 8.4742\\
\hline\hline
\end{tabular}
\end{center}
\end{table}

The ground state energy of H atom in these three representations are respectively -0.991 Ry, -0.99912 Ry and -0.99987 Ry. Additionally, the expressions (11) and (12), can be used to evaluate the hopping integral $<i|H_i|j>$.

The main reason for introducing the Gaussian is to be able to calculate efficiently the many-site integrals involving the repulsive Coulomb interactions, which appear in evaluation of $U$ and $K$ parameters of the parameterized Hamiltonian (1) in Sec. 2. In general the Coulomb integral $<ij|V|kl>=V_{ijkl}$ has the following form in the Gaussian representation
\begin{equation}
V_{ijkl}^{abcd}  \equiv \left\langle {g_i^{(a)} g_j^{(b)} } \right|V\left| {g_k^{(c)} g_l^{(d)} } \right\rangle  =\end{equation}
$$
S_{ac} (R_{ik} )S_{bd} (R_{jl} )\frac{2}{{R_{ijkl} }}{\rm{erf}}\left( {b_{abcd}^2 R_{ijkl}^{abcd} /B_{abcd} } \right)
$$
where
\begin
{equation}R_{ijkl}^{abcd}  \equiv \left| {\left( {{\bf{R}}_i  + \frac{{{\bf{R}}_{ik} }}{{1 + \left( {\Gamma _a /\Gamma _c } \right)^2 }}} \right) - \left( {{\bf{R}}_j  + \frac{{{\bf{R}}_{jl} }}{{1 + \left( {\Gamma _b /\Gamma _d } \right)^2 }}} \right)} \right|
\end{equation}
\begin{equation}
b_{abcd}  = \left( {\frac{{\Gamma _a^2  + \Gamma _c^2 }}{2}} \right)^{1/2} \left( {\frac{{\Gamma _b^2  + \Gamma _d^2 }}{2}} \right)^{1/2},
\end{equation}
\begin{equation}
B_{abcd}  = \left( {\frac{{\Gamma _a^2  + \Gamma _b^2  + \Gamma _c^2  + \Gamma _d^2 }}{4}} \right)^{1/2}.
\end{equation}
Using those expression, the following expressions for parameters $U'$ and $K'$ (i.e. those calculated for the atomic wave functions; $U$ and $K$ are those calculated in the Wannier representation \cite{SpalekPodsiadly}:

\begin{equation}
U' \equiv \left\langle {\Psi _i \Psi _i } \right|V\left| {\Psi _i \Psi _i } \right\rangle  = 
\end{equation}
$$\alpha ^6 \sum\limits_{a,b,c,d} {\beta _a \beta _b \beta _c \beta _d \frac{{32}}{{\pi ^{1/2} }}\left( {\frac{{\Gamma _a \Gamma _b \Gamma _c \Gamma _d }}{{\left( {\Gamma _a^2  + \Gamma _c^2 } \right)\left( {\Gamma _b^2  + \Gamma _d^2 } \right)}}} \right)^{3/2} \frac{{b_{abcd}^2 }}{{B_{abcd} }}}
$$
and
\begin{equation}
K' \equiv \left\langle {\Psi _i \Psi _j } \right|V\left| {\Psi _i \Psi _j } \right\rangle  =
\end{equation}
$$\alpha ^6 \sum\limits_{a,b,c,d} {\beta _a \beta _b \beta _c \beta _d \frac{2}{{R_{ij} }}{\rm{erf}}\left( {b_{abcd}^2 R_{ij} /B_{abcd} } \right)}.
$$

Explicit calculation of the above integrals requires a number of numerical operations. For example evaluation of integral (22) involves computation of 81 integrals in STO-3G basis and 2401 integrals in STO-7G basis. Nevertheless, they can be calculated efficiently. Obviously, the whole procedure still requires the minimization of the total state energy with respect to $\alpha$ in the correlated state.

\end{document}